\documentclass[12pt]{article}
\usepackage{amsfonts}
\usepackage{graphicx}
\usepackage{amssymb}
\usepackage{amscd}
\usepackage{amsmath}
\begin{document}
\renewcommand{\thefootnote}{\fnsymbol{footnote}}
\begin{titlepage}
%\begin{flushright}
%arXiv:0912.xxxx[hep-th]
%\end{flushright}

\vspace{10mm}
\begin{center}
{\Large\bf Investigation of $PT$-symmetric Hamiltonian systems from an alternative point of view}
\vspace{16mm}

{\large Jun-Qing Li, Qian Li, and
Yan-Gang Miao\footnote{Corresponding author. E-mail: miaoyg@nankai.edu.cn}

\vspace{6mm}
{\normalsize \em School of Physics, Nankai University, Tianjin 300071, China}}

\end{center}

\vspace{10mm}
\centerline{{\bf{Abstract}}}
\vspace{6mm}
\noindent
Two non-Hermitian $PT$-symmetric Hamiltonian systems are reconsidered by means of the algebraic method
which was originally proposed for the pseudo-Hermitian Hamiltonian systems rather than for the $PT$-symmetric ones.
Compared with the way converting a non-Hermitian Hamiltonian to its Hermitian counterpart,
this method has the merit that keeps the Hilbert space of the non-Hermitian $PT$-symmetric Hamiltonian unchanged.
In order to give the positive definite inner product for the $PT$-symmetric systems, a new operator $V$, instead of $C$, can be introduced.
The operator $V$  has the similar function to the operator $C$ adopted normally in the $PT$-symmetric quantum mechanics, however,
it can be constructed, as an advantage, directly in terms of Hamiltonians. The spectra of the two non-Hermitian $PT$-symmetric systems
are obtained,  which coincide with that given in literature, and in particular,
the Hilbert spaces associated with positive definite inner products are worked out.

\vskip 20pt
\noindent
{\bf PACS Number(s)}: 11.30.Er, 03.65.-w, 03.65.Fd, 03.65.Ge

\vskip 20pt
\noindent
{\bf Keywords}: $PT$ symmetry, positive definite inner product, algebraic method

\end{titlepage}
\newpage
\renewcommand{\thefootnote}{\arabic{footnote}}
\setcounter{footnote}{0}
\setcounter{page}{2}

\section{Introduction}

One class of non-Hermitian Hamiltonians has positive and real spectra~\cite{BBR,BBMP} if the non-Hermitian Hamiltonian satisfies the condition:
both the Hamiltonian and its eigenfunctions are $PT$ invariant, where the linear parity operator $P$ reverses the
position and momentum: $x\rightarrow -x,\ p\rightarrow -p$, and the antilinear time reversal operator $T$ reverses
the momentum and
imaginary unit: $p\rightarrow -p, \ i\rightarrow -i$.
If the inner product of two states $\varphi(x)$ and $\phi(x)$ is defined~\cite{BQZG} to be:
$\langle\varphi(x),\phi(x)\rangle_{PT}\equiv\int [PT \varphi(x)]\phi(x) dx$, such a quantity is not positive definite.
This problem has been overcome~\cite{BBJM} by introducing the operator $C$ that commutes  with both the non-Hermitian $PT$-symmetric
Hamiltonian and the combined operator
$PT$. That is to say, the $CPT$ inner product turns out to be positive definite, the Hamiltonian and its transposition are related~\cite{BBJM} by the
$CPT$ similarity transformation in addition to the Hamiltonian's $PT$ symmetry,
and the time evolution generated by the Hamiltonian is kept unitary in the $PT$-symmetric theory. Therefore,
the eigenfunctions of the non-Hermitian $PT$-symmetric Hamiltonian can be orthogonal and complete~\cite{SW}
such that the non-Hermitian $PT$-symmetric theory can have a probability interpretation.
In addition,
the breaking of the $PT$ symmetry has been observed in experiments~\cite{LF} in the realm of optics.
Consequently,
the basic frame of
the non-Hermitian $PT$-symmetric quantum mechanics has been established.

Another class of non-Hermitian quantum theory that has been studied recently is closely related to a
pseudo-Hermitian (or quasi-Hermitian) Hamiltonian~\cite{Pauli,SGH,Ali1,Ali2,Ali3}.
The pseudo-Hermitian theory with an indefinite metric operator $\eta$ was first proposed
by Pauli~\cite{Pauli} in 1943 for the sake of overcoming the divergence of quantum field theories. Later, the theory
with a positive definite metric $\eta_+$ was developed by others~\cite{SGH,Ali1,Ali2,Ali3}. % in order to give positive definite inner products.
That is, the Hamiltonian is
$\eta_+$ pseudo-Hermitian self-adjoint and its eigenfunctions have positive definite inner products with respect to this positive definite metric.
As an important progress of the two classes of
the non-Hermitian quantum mechanics, there is an intimate relation~\cite{ME} between the $PT$-symmetric and
the pseudo-Hermitian Hamiltonians, i.e.,
an exact antilinear symmetric system\footnote{The $PT$ symmetry is only a specific case of the antilinear symmetry.}
can be transformed
into its corresponding Hermitian system through a similarity transformation, where the similarity transformation can be realized by means of
the indefinite metric operator $\eta$.
Specifically, a $PT$-symmetric Hamiltonian can correspond to a Hermitian one.
The method of transforming a non-Hermitian system into a Hermitian one has frequently been used recently,
such as dealing with the fourth-order derivative Pais-Uhlenbeck oscillator model~\cite{BMN} and the $PT$-symmetric Hamiltonian systems that
are composed of interacting non-Hermitian and Hermitian Hamiltonians~\cite{BJI}.

However, we have to point out that the method used in ref.~\cite{BJI}, i.e., converting a non-Hermitian $PT$-symmetric model to
its corresponding Hermitian one by means of a similarity transformation,
alters the Hilbert space of the non-Hermitian Hamiltonian system. In other words, one can easily verify that
the commutator of the two Hamiltonians (the non-Hermitian Hamiltonian and its Hermitian counterpart) is non-vanishing,
which means that they give rise to different Hilbert spaces,
and that the similarity transformation only ensures the same spectrum for the two Hamiltonians.
To this end, we need an improved method which is available for one to get the spectrum of the non-Hermitian $PT$-symmetric
Hamiltonian but not to alter the Hilbert space spanned
by the eigenfunctions of the Hamiltonian.
Actually, such a method, called the algebraic method, has already been proposed~\cite{LMX} for the pseudo-Hermitian Hamiltonian
systems~\cite{Pauli,SGH,Ali1,Ali2,Ali3}.
Here we find that this algebraic method is also available to the non-Hermitian $PT$-symmetric Hamiltonian systems analyzed in detail in ref.~\cite{BJI}.
That is, we shall investigate in terms of the algebraic method
the two non-Hermitian $PT$-symmetric Hamiltonian systems considered in ref.~\cite{BJI},
and achieve the goal that the same spectra as that given in ref.~\cite{BJI} are obtained and further the Hilbert spaces
with positive definite inner products are worked out. We note that it is a key step to construct the operator $V$ that is model dependent
like the operator $C$.
The operator $V$ has the similar function to the operator $C$ adopted~\cite{BBJM} normally in the $PT$-symmetric quantum mechanics, that is, to make
the inner product positive definite. The reason for us to introduce $V$  is that it
can be constructed directly in terms of the Hamiltonians of quantum systems, and thus the formulation of $V$ is more intuitive than that of $C$.

This paper is organized as follows. In the next section,
we first diagonalize the non-Hermitian $PT$-symmetric Hamiltonian~\cite{BJI} which is composed of
two coupled $PT$-symmetric Hamiltonians, where one is Hermitian and the other non-Hermitian. Then, we construct the operator $V$ which ensures that
the $PT$-symmetric system now has a positive definite inner product with respect to the combined operator $PTV$. Next,
we redefine the annihilation and creation operators for the $PT$-symmetric Hamiltonian and give the real spectrum with a lower bound
by means of the algebraic method. The spectrum we obtain is the same as that given in ref.~\cite{BJI}, and in particular,
we provide the complete set of eigenfunctions with the
positive definite $PTV$ inner product.
In section 3, we extend our investigation to a more complicated model composed of two coupled non-Hermitian $PT$
symmetric Hamiltonians and fulfill the task similar to that of section 2.
Finally, section 4 is devoted to a brief conclusion.

\section{Model 1: A coupled Hermitian and non-Hermitian $PT$-symmetric Hamiltonian system}

We deal with the model given in ref.~\cite{BJI} by means of the algebraic method~\cite{LMX}. The Hamiltonian of the model takes the form,
\begin{equation}
H=\left(p^2_1+x^2_1\right)+\left(p^2_2+x^2_2+i2x_2\right)+2\epsilon x_1 x_2,\label{PTH}
\end{equation}
which is composed of the (Hermitian) harmonic oscillator Hamiltonian, the non-Hermitian $PT$-symmetric Hamiltonian and the interacting
Hamiltonian with the coupling constant\footnote{The energy spectrum is real and positive when $|\epsilon| < 1$. The critical value is at
$|\epsilon| = 1$, and the spectrum becomes complex when $|\epsilon| > 1$. For the
details, see ref.~\cite{BJI}. In the present paper we focus first on the region of $|\epsilon| < 1$, and then point out particularly that the reason that
the complex spectrum occurs  if $|\epsilon| > 1$ is just the breaking of the $PT$ symmetry, which was unanswered in
ref.~\cite{BJI} because the eigenfunctions of the Hamiltonian system were not obtained there.}
$\epsilon$.
Note that $(x_j, p_j)$, where $j=1,2$, are two pairs of canonical variables that satisfy the usual commutation relations,
\begin{equation}
[x_j, p_k]=i{\delta}_{jk}, \qquad [x_j, x_k]=0=[p_j ,p_k], \qquad
j,k=1,2,\label{cr1}
\end{equation}
where $\hbar$ is set to be unity throughout this paper.

As pointed out in the above section,
the Hamiltonian eq.~(\ref{PTH}) was converted to the following Hermitian one  by a similarity transformation in ref.~\cite{BJI},
\begin{equation}
\mathbf{H}=\left(p^2_1+x^2_1\right)+\left(p^2_2+x^2_2\right)+2\epsilon x_1 x_2+\frac{1}{1-\epsilon^2}.\label{HH}
\end{equation}
It is easy to check that $[H, \mathbf{H}]=-4p_2 \neq 0$, which means that $H$ and $\mathbf{H}$ have different sets of
eigenfunctions, i.e., they give different Hilbert spaces.
Thus $H$ and $\mathbf{H}$ describe two different systems although they have the same spectrum.
In order to find out the Hilbert space with the positive definite inner products for the non-Hermitian system $H$, we turn to the use
of the algebraic method~\cite{LMX} which has been proved to be available in dealing with the pseudo-Hermitian systems.

\subsection{Diagonalization}

Let us diagonalize the Hamiltonian eq.~(\ref{PTH}). Applying the way used in ref.~\cite{BJI} directly to this non-Hermitian Hamiltonian,
we introduce the new variables of phase space $(X_j, P_j)$, where $j=1,2$, and establish the relations between the new and original variables as follows,
\begin{eqnarray}
p_1=aP_1+bP_2,\quad p_2=cP_1+dP_2,\quad x_1=eX_1+fX_2,\quad x_2=gX_1+hX_2,\label{replacement}
\end{eqnarray}
where $a, b, c, d, e, f, g$, and $h$ are unknown real coefficients. Furthermore, we impose the canonical commutation relations as eq.~(\ref{cr1})
to the new variables,
\begin{equation}
[X_j,P_k]=i{\delta}_{jk}, \qquad [X_j,X_k]=0=[P_j,P_k], \qquad
j,k=1,2.\label{cr2}
\end{equation}
We determine the unknown coefficients by requiring  that (i) The canonical commutation relations for the new variables and for the original ones
should be consistent to each other, and (ii) The cross terms of the Hamiltonian expressed in terms of the new variables $(X_j, P_j)$
should be vanished when eq.~(\ref{replacement}) is substituted into eq.~(\ref{PTH}). The two sets of conditions give six equations for the eight
unknown coefficients, and the solutions are
\begin{eqnarray}
c=\zeta a,\qquad d=-\zeta b,\qquad e=\frac{1}{2a},\qquad f=\frac{1}{2b},\qquad g=\frac{\zeta}{2a},\qquad
h=-\frac{\zeta}{2b},\label{covalue}
\end{eqnarray}
where $\zeta=\pm 1$, and $a$ and $b$ are arbitrary non-vanishing real parameters. As a result, the Hamiltonian eq.~(\ref{PTH}) is now diagonalized as
\begin{eqnarray}
H=2a^2P^2_1+\frac{1+\zeta\epsilon}{2a^2}\left(X_1+i\frac{a\zeta}{1+\zeta\epsilon}\right)^2+2b^2P^2_2+
\frac{1-\zeta\epsilon}{2b^2}\left(X_2-i\frac{b\zeta}{1-\zeta\epsilon}\right)^2+\frac{1}{1-\epsilon^2},\label{diagH}
\end{eqnarray}
and the relations between the new and original variables are given by
\begin{eqnarray}
P_1=\frac{1}{2a}\left(p_1+\zeta p_2\right), && P_2=\frac{1}{2b}\left(p_1-\zeta p_2\right),\nonumber\\
X_1=a\left(x_1+\zeta x_2\right), && X_2=b\left(x_1-\zeta x_2\right).\label{arguments}
\end{eqnarray}

In order to utilize the algebraic method conveniently, we further introduce two pairs of variables in phase space,
$({\cal{X}}_j, {\cal{P}}_j)$, where $j=1,2$, and rewrite the above Hamiltonian (eq.~(\ref{diagH})) in a completely
diagonalized form,
\begin{eqnarray}
H&=&H_1+H_2+\frac{1}{1-\epsilon^2},\nonumber\\
H_1&=&2a^2{\cal{P}}_1^2+\frac{1+\zeta\epsilon}{2a^2}{\cal{X}}_1^2, \nonumber\\
H_2&=&2b^2{\cal{P}}_2^2 +\frac{1-\zeta\epsilon}{2b^2}{\cal{X}}_2^2,\label{comdiagH}
\end{eqnarray}
where ${\cal{X}}_j$ and  ${\cal{P}}_j$ are defined as follows:
\begin{eqnarray}
{\cal{P}}_1:=P_1, && {\cal{P}}_2:=P_2,\nonumber\\
{\cal{X}}_1:=X_1+i\frac{a\zeta}{1+\zeta\epsilon}, && {\cal{X}}_2:=X_2-i\frac{b\zeta}{1-\zeta\epsilon}.\label{Arguments}
\end{eqnarray}
We emphasize that ${{\cal{P}}_j}$'s are still Hermitian while ${\cal{X}}_j$'s non-Hermitian due to the non-Hermiticity of the Hamiltonian eq.~(\ref{PTH}),
which is different from the case occurred in ref.~\cite{BJI} but suitable for being dealt with by the algebraic method~\cite{LMX}.
Note that $({\cal{X}}_j, {\cal{P}}_j)$ satisfy the same commutation relations as
$(X_j, P_j)$,
\begin{equation}
[{\cal{X}}_j, {\cal{P}}_k]=i{\delta}_{jk}, \qquad [{\cal{X}}_j, {\cal{X}}_k]=0=[{\cal{P}}_j, {\cal{P}}_k], \qquad
j,k=1,2,\label{cr3}
\end{equation}
which meet the basic requirement for us to apply the algebraic method to the Hamiltonian system described by
eqs.~(\ref{comdiagH}) and (\ref{Arguments}).

\subsection{$CPT$ inner product and its shortcoming}

We deviate our goal temporarily and mention the normally used $CPT$ inner product and its shortcoming,
which may provide some reason  for us to adopt our $PTV$ inner product (or its equivalent $PV$-pseudo inner product) in the next subsection.
As analyzed in refs.~\cite{BBJM,SW},
the inner product of eigenfunctions in $PT$-symmetric Hamiltonian systems is not positive definite,
\begin{equation}
\langle\varphi_n(x),\varphi_m(x)\rangle_{PT}\equiv\int[PT\varphi_n(x)]\varphi_m(x)dx=(-1)^n{\delta}_{nm},\label{norm1}
\end{equation}
where $\{\varphi_n(x), \, n \subset \mathbb{N}\}$ is the set of eigenfunctions of a $PT$-symmetric Hamiltonian.
In order to overcome this difficulty, a linear operator
$C$ is constructed~\cite{BBJM} in terms of the set of eigenfunctions in such a way that it
commutes with both the $PT$-symmetric Hamiltonian and the combined operator $PT$, and in particular that
it has the following desired property,
\begin{equation}
C \varphi_n(x)=(-1)^n\varphi_n(x),\label{functC}
\end{equation}
where the property $C^2=1$ is obvious.
Consequently, the $CPT$ inner product turns out to be positive definite,
\begin{equation}
\langle\varphi_n(x),\varphi_m(x)\rangle_{CPT}\equiv\int[CPT\varphi_n(x)]\varphi_m(x)dx={\delta}_{nm}.\label{norm2}
\end{equation}

However, the operator $C$ is unknown before the eigenfunctions of a non-Hermitian $PT$-symmetric Hamiltonian system are solved, and it is
hard to be expressed concisely even after the eigenfunctions are obtained. This reminds us to search for an alternative
operator which, associated directly with the non-Hermitian $PT$-symmetric Hamiltonian rather than its eigenfunctions,
not only maintains the desired property (eq.~(\ref{functC})) but also is easy to be constructed.
Fortunately, such a substitutor can be found out. See the definition of the operator $V$ in the next subsection.

\subsection{$PTV$ inner product and its advantage}

Now we turn to our $PTV$ inner product (or its equivalent $PV$-pseudo inner product)
in this subsection and then give in the next subsection the energy spectrum and eigenfunctions
for the non-Hermitian $PT$-symmetric system described by the Hamiltonian eq.~(\ref{PTH}) or eq.~(\ref{comdiagH}).

According to the $PT$-symmetric quantum mechanics with the positive definite $CPT$ inner product~\cite{BBJM}, a Hamiltonian $\mathcal{H}$,
in addition to the $PT$ symmetry $\mathcal{H}=(PT)^{-1}\mathcal{H}(PT)$, is required to satisfy
\begin{equation}
\mathcal{H}=(CPT)^{-1}\tilde{\mathcal{H}}(CPT),\label{CPTsym}
\end{equation}
where the tilde stands for transposition. Considering the properties $[P, T]=0$ and $[C, PT]=0$ given in ref.~\cite{BBJM}, we
can reduce eq.~(\ref{CPTsym}) to
\begin{equation}
\mathcal{H}=(TPC)^{-1}\tilde{\mathcal{H}}(TPC)=(PC)^{-1}\cdot T^{-1}\tilde{\mathcal{H}}T\cdot (PC)=(PC)^{-1}\mathcal{H}^{\dag}(PC),\label{CPTsym2}
\end{equation}
where the dagger means Hermitian conjugate. We note that eq.~(\ref{CPTsym2}) establishes a relationship between a non-Hermitian
$PT$-symmetric quantum system and a $PC$ pseudo-Hermitian one~\cite{Pauli}.
That is, the requirement that a $PT$-symmetric Hamiltonian has a positive definite
$CPT$ inner product leads to the result\footnote{Although it has been mentioned in ref.~\cite{Ali3}, this result is obtained here with no use of
the postulation that $\mathcal{H}$ has a complete biorthonormal eigenbasis and a discrete spectrum.
We note that this postulation that is not mandatory here is crucial for the outcomes deduced in ref.~\cite{Ali3}.}
that this $PT$-symmetric Hamiltonian must be $PC$
pseudo-Hermitian self-adjoint.
Consequently, we can bring the $PT$-symmetric Hamiltonian system into the framework of the $PC$ pseudo-Hermitian Hamiltonian system.
Different from ref.~\cite{Ali3}, our proof is achieved at the
level of Hamiltonians, which is not involved in eigenfunctions that are hard to be solved sometimes (see also footnote 3).

The key step of the algebraic method is to work out the operator $V$ that has the similar function to that of $C$ but is only associated with the
Hamiltonian of the system we are investigating, while the operator  $C$
is relevant to the eigenfunctions of the Hamiltonian. The operator $V$ should be constructed~\cite{LMX} in terms of the Hamiltonian
(see eqs.~(\ref{comdiagH}) and (\ref{Arguments}))
in the following way,
\begin{eqnarray}
V = e^{i\pi\left(\frac{H_1}{2\sqrt{1+\zeta\epsilon}}
+\frac{H_2}{2\sqrt{1-\zeta\epsilon}}-1\right)}. \label{V}
\end{eqnarray}
On the one hand, from the point of view of the $PT$-symmetric quantum mechanics, one can verify
that $V$ indeed has the same properties\footnote{The property $V^2=1$ will be shown after $V$'s eigenfunctions
that are also the eigenfunctions of the Hamiltonian (eq.~(\ref{PTH}) or eq.~(\ref{comdiagH})) are solved.}  as $C$.
That is,  $V$ is non-Hermitian but $PT$-symmetric, and it commutes with both $PT$ and the Hamiltonian (eq.~(\ref{PTH}) or eq.~(\ref{comdiagH})).
On the other hand, from the point of view of the pseudo-Hermitian quantum mechanics, one can verify that $V$ is $P$-pseudo-Hermitian self-adjoint
because the Hamiltonian (eq.~(\ref{PTH}) or eq.~(\ref{comdiagH})) is $P$-pseudo-Hermitian self-adjoint,
i.e., $V=P^{-1}V^{\dag}P$ due to $H=P^{-1}H^{\dag}P$, the positive definite metric operator defined by $\eta_+ := PV$ is Hermitian as desired, i.e.,
$\eta_+^\dagger=V^\dagger P^\dagger=V^\dagger P=P (P^{-1}V^\dagger P)=P V=\eta_+$, and therefore
the Hamiltonian (eq.~(\ref{PTH}) or eq.~(\ref{comdiagH}))
is further $\eta_+$-pseudo-Hermitian self-adjoint, i.e., $H={\eta_+}^{-1}H^{\dag}\eta_+=(PV)^{-1}H^{\dag}(PV)$. In particular, the pseudo inner product
associated with the positive definite metric operator $\eta_+$ can be shown\footnote{In the $PT$-symmetric quantum mechanics, the $PTV$ inner product
is defined as $\langle\varphi(x),\phi(x)\rangle_{PTV}\equiv\int[PTV\varphi(x)]\phi(x)dx$, while in the $PV$-pseudo-Hermitian quantum
mechanics, the $PV$-pseudo inner product is defined as
$\langle\varphi(x)|\phi(x)\rangle_{PV}\equiv\int \overline{\varphi}(x) PV \phi(x)dx$, where the overline stands for complex
conjugate. By using the properties: $[P, T]=0$, $[V, PT]=0$, and $(PV)^{\dag}=PV$, one can prove that the two definitions of inner products
are equivalent, i.e., $\langle\varphi(x),\phi(x)\rangle_{PTV}=\langle\varphi(x)|\phi(x)\rangle_{PV}$.}
to be equivalent to the $PTV$ inner product and
thus it is positive definite. The advantage of using $V$ instead of $C$ is not only that $V$ can easily be constructed but also that
we are enlightened to apply the algebraic method (that was originally proposed for the pseudo-Hermitian quantum mechanics)
to the $PT$-symmetric quantum mechanics.

By means of the operator $V$ given above, we can now define new annihilation and creation operators by following the algebraic method.
When the annihilation operators take the forms,
\begin{eqnarray}
a_1 &=& \frac{a}{\sqrt[4]{1+\zeta\epsilon}}\left(i{\cal{P}}_1+\frac{\sqrt{1+\zeta\epsilon}}{2a^2}{\cal{X}}_1\right), \nonumber \\
a_2&=&\frac{b}{\sqrt[4]{1-\zeta\epsilon}}\left(i{\cal{P}}_2+\frac{\sqrt{1-\zeta\epsilon}}{2b^2}{\cal{X}}_2\right),\label{cran}
\end{eqnarray}
the corresponding creation operators are defined as the $PV$-pseudo Hermitian adjoint of the annihilation
operators,\footnote{Within the framework of the $PT$-symmetric quantum mechanics, the creation operators can also be expressed as
$a^{\ddagger}_1=(PTV)^{-1}\tilde{a}_1(PTV)$ and $a^{\ddagger}_2=(PTV)^{-1}\tilde{a}_2(PTV)$. We can verify them easily by referring to
eq.~(\ref{CPTsym2}).}
\begin{eqnarray}
a^{\ddagger}_1\equiv(PV)^{-1}a_1^{\dag}(PV)=\frac{a}{\sqrt[4]{1+\zeta\epsilon}}\left(-i{\cal{P}}_1+\frac{\sqrt{1+\zeta\epsilon}}{2a^2}{\cal{X}}_1\right),
\nonumber \\
a^{\ddagger}_2\equiv(PV)^{-1}a_2^{\dag}(PV)=\frac{b}{\sqrt[4]{1-\zeta\epsilon}}\left(-i{\cal{P}}_2+\frac{\sqrt{1-\zeta\epsilon}}{2b^2}{\cal{X}}_2\right).
\label{cran2}
\end{eqnarray}
Using the commutation relations eq.~(\ref{cr3}) satisfied by the variables of phase space $({\cal{X}}_j,{\cal{P}}_j)$, where $j=1,2$,
we can verify that the newly defined annihilation and creation operators  satisfy the expected algebraic relations
\begin{eqnarray}
[a_j,a^{\ddag}_k]=\delta_{jk}, \qquad
[a_j,a_k]=0=[a^{\ddag}_j,a^{\ddag}_k], \qquad  j,k=1,2.\label{acr}
\end{eqnarray}

We define the number operator\footnote{Repeated subscripts do not sum except for extra indications.}
in the similar way to that in the conventional quantum mechanics,
\begin{eqnarray}
N_j=a^{\ddagger}_j a_j, \qquad j=1,2,\label{number}
\end{eqnarray}
and obtain the other expected algebraic relations by using eqs.~(\ref{acr}) and (\ref{number}),
\begin{eqnarray}
[N_j,a^{\ddagger}_k]=a^{\ddagger}_j\delta_{jk}, \qquad [N_j,a_k]=-a_j\delta_{jk},\qquad j,k=1,2.\label{Nacr2}
\end{eqnarray}
Furthermore, for a given set of eigenstates of the number operator $N_j$, i.e., $|n_j\rangle$ , we have
\begin{eqnarray}
N_j|n_j\rangle=n_j|n_j\rangle, \qquad j=1,2.\label{eigenstate}
\end{eqnarray}
When considering the equivalence between the $PTV$ inner product in the $PT$-symmetric quantum mechanics and the $PV$-pseudo inner product in
the $PV$-pseudo Hermitian quantum, and utilizing the positive definiteness of the two classes of inner products (see eq.~(\ref{norm2}) and footnote 5),
we finally convince that the operators $a_j$ and $a^{\ddagger}_j$  (see eqs.~(\ref{cran}) and (\ref{cran2}))
are indeed annihilation and creation operators, respectively, and have the property of ladder operators,
\begin{eqnarray}
a^{\ddagger}_j|n_j\rangle = \sqrt{n_j+1}\,|n_j+1\rangle,\qquad a_j|n_j\rangle =\sqrt{n_j}\,|n_j-1\rangle, \qquad j=1,2.\label{ladder1}
\end{eqnarray}
Consequently, we rewrite the $PT$-symmetric Hamiltonian (see eqs.~(\ref{comdiagH}) and (\ref{Arguments})) in terms of the number operators as follows:
\begin{eqnarray}
H=\sqrt{1+\zeta\epsilon}\left(2N_1+1\right)+\sqrt{1-\zeta\epsilon}\left(2N_2+1\right)+\frac{1}{1-\epsilon^2}.\label{numberH}
\end{eqnarray}

\subsection{Spectrum and eigenfunction}

Now we obtain the energy spectrum from eq.~(\ref{numberH}),
\begin{equation}
E_{n_1n_2}=\sqrt{1+\zeta\epsilon}\left(2n_1+1\right)+\sqrt{1-\zeta\epsilon}\left(2n_2+1\right)+\frac{1}{1-\epsilon^2},\label{spectrum}
\end{equation}
where $n_1, n_2 \subset \mathbb{N}$.
This result is obviously same as that given in ref.~\cite{BJI} but here it is derived in terms of the algebraic method, which shows that
the algebraic method is also available for non-Hermitian $PT$-symmetric quantum systems.

Next, we focus on the eigenfunctions of the system described by the Hamiltonian eq.~(\ref{comdiagH})  or
eq.~(\ref{numberH}), which is beyond the context of ref.~\cite{BJI}.
Quite similar to the case of the two separate harmonic oscillators, we solve 
$H\varphi_{n_1n_2}({\cal{X}}_1,{\cal{X}}_2)=E_{n_1n_2}\,\varphi_{n_1n_2}({\cal{X}}_1,{\cal{X}}_2)$ and obtain the  
eigenfunctions with the help of the Mathematica,
 \begin{eqnarray}
\varphi_{n_1n_2}({\cal{X}}_1,{\cal{X}}_2)=\varphi_{n_1}({\cal{X}}_1)\,\varphi_{n_2}({\cal{X}}_2),\label{eigenstate}
\end{eqnarray}
where  ${\cal{X}}_1$ and ${\cal{X}}_2$ are now denoted as the coordinates whose operators are defined in eq.~(\ref{Arguments}),
and the eigenfunctions of the ``single harmonic oscillator"  take the form,
\begin{equation}
\varphi_{n_j}({\cal{X}}_j)=\frac{\sqrt{c_j}}{\sqrt[4]{\pi}}(2^{n_j}n_j!)^{-\frac{1}{2}}e^{-\frac{1}{2}(c_j {\cal{X}}_j)^2}H_{n_j}(c_j{\cal{X}}_j),
\qquad j=1,2. \label{state}
\end{equation}
Note that $H_{n_j}(c_j{\cal{X}}_j)$ is the Hermite polynomial of the $n_j$-th degree, where
$c_j$'s are parameters given by
\begin{equation}
c_1=\frac{\sqrt[4]{1+\zeta\epsilon}}{\sqrt{2a^2}}, \qquad  c_2=\frac{\sqrt[4]{1-\zeta\epsilon}}{\sqrt{2b^2}},\label{c1c2}
\end{equation}
which are real when $|\epsilon| < 1$.

At this stage we can complete the proof of the property $V^2=1$ for the operator $V$ (see eq.~(\ref{V})) and the positive
definiteness of inner products.
As $V$ commutes with the Hamiltonian (see eq.~(\ref{comdiagH}) or eq.~(\ref{numberH})),
$\varphi_{n_1n_2}({\cal{X}}_1,{\cal{X}}_2)$ is also the set of eigenfunctions of $V$. Therefore,
by using eqs.~(\ref{V}), (\ref{eigenstate}), and (\ref{state}) we get
\begin{eqnarray}
V\varphi_{n_1n_2}({\cal{X}}_1,{\cal{X}}_2)=(-1)^{n_1+n_2}\varphi_{n_1n_2}({\cal{X}}_1,{\cal{X}}_2), \label{VV}
\end{eqnarray}
which gives rise to the expected property $V^2=1$. Furthermore,
the above equation coincides with that of the operator $C$ (see eq.~(\ref{functC}))
used for constructing the positive definite inner product in $PT$-symmetric systems.
As analyzed in ref.~\cite{LMX}, by using the
Cauchy's residue theorem, the properties of the Hermite polynomials and eqs.~(\ref{eigenstate})-(\ref{VV}), we can verify that
the $PV$-pseudo inner product of the eigenfunctions is positive definite and
orthogonal\footnote{Equivalently, it can be expressed in the notations of the
$PT$-symmetric quantum mechanics as the $PTV$ inner product (cf. footnote 5.):
\begin{eqnarray}
& & \langle\varphi_{n_1n_2}({\cal{X}}_1,{\cal{X}}_2)|\varphi_{m_1m_2}({\cal{X}}_1,{\cal{X}}_2)\rangle_{PV} \nonumber \\
&=&\langle\varphi_{n_1n_2}({\cal{X}}_1,{\cal{X}}_2),\varphi_{m_1m_2}({\cal{X}}_1,{\cal{X}}_2)\rangle_{PTV}
\nonumber \\
&\equiv &\int_{-\infty+iI_1}^{+\infty+iI_1}\int_{-\infty+iI_2}^{+\infty+iI_2}
\left[PTV{\varphi}_{n_1n_2}({\cal{X}}_1,{\cal{X}}_2)\right]  \varphi_{m_1m_2}({\cal{X}}_1,{\cal{X}}_2) d{\cal{X}}_1 d{\cal{X}}_2 \nonumber \\
&=&{\delta}_{n_1m_1}{\delta}_{n_2m_2}.\nonumber
\end{eqnarray}},
\begin{eqnarray}
& & \langle\varphi_{n_1n_2}({\cal{X}}_1,{\cal{X}}_2)|\varphi_{m_1m_2}({\cal{X}}_1,{\cal{X}}_2)\rangle_{PV}
\nonumber \\
&\equiv &\int_{-\infty+iI_1}^{+\infty+iI_1}\int_{-\infty+iI_2}^{+\infty+iI_2}
\overline{\varphi}_{n_1n_2}({\cal{X}}_1,{\cal{X}}_2) PV \varphi_{m_1m_2}({\cal{X}}_1,{\cal{X}}_2) d{\cal{X}}_1 d{\cal{X}}_2 \nonumber \\
&=&{\delta}_{n_1m_1}{\delta}_{n_2m_2},
\label{pnorm}
\end{eqnarray}
where $I_1$ and $I_2$ are two real parameters which can be determined from eq.~(\ref{Arguments}), i.e.,
$I_1=\frac{a\zeta}{1+\zeta\epsilon}$ and $I_2=-\frac{b\zeta}{1-\zeta\epsilon}$.

\subsection{Breaking of the $PT$ symmetry}
In the above subsections we focus only on the region of  $|\epsilon| < 1$ in which the energy spectrum is real and positive,
see eq.~(\ref{spectrum}), and the eigenfunctions of the Hamiltonian, see eqs.~(\ref{eigenstate}) and (\ref{state}),
 are also the eigenfunctions of the operator $PT$,
which can be seen clearly from the following equation,
\begin{eqnarray}
PT \varphi_{n_1n_2}({\cal{X}}_1,{\cal{X}}_2)=(-1)^{n_1+n_2}\varphi_{n_1n_2}({\cal{X}}_1,{\cal{X}}_2), \label{PTsymmetric}
\end{eqnarray}
i.e., the $PT$ symmetry of the Hamiltonian system is unbroken.

Note that $|\epsilon| = 1$ is a critical point at which the model described by eq.~(\ref{PTH}) is no longer a free two-dimensional
oscillator-like system. This can be verified after eq.~(\ref{PTH}) under this critical condition is
diagonalized. %Thus the critical case becomes meaningless.
When $|\epsilon| > 1$, one can see obviously from eq.~(\ref{spectrum}) that the spectrum
becomes complex, where the parameter $\zeta$ takes $1$ or $-1$.
We note that the Hamiltonian depicted by eq.~(\ref{PTH}) or eq.~(\ref{comdiagH}) is $PT$ symmetric, which is independent of
the magnitude of the coupling constant $\epsilon$. In the region of $|\epsilon| > 1$, eq.~(\ref{eigenstate}) is still the set of eigenfunctions of
the Hamiltonian but no longer that of the operator $PT$, i.e., the $PT$ symmetry is broken.
Let us verify this result. As either $c_1$ or
$c_2$ turns out to be complex if $|\epsilon| > 1$, see eq.~(\ref{c1c2}), that is,
each of the two cases must happen, we take the case of a complex $c_1$ and a real $c_2$ as an example which corresponds to (i) $\zeta =1$ and
$\epsilon < -1 $ or (ii) $\zeta = -1$ and
$\epsilon > 1 $.
In this case, $\varphi_{n_1}({\cal{X}}_1)$ is not the eigenfunction of $PT$ although $\varphi_{n_2}({\cal{X}}_2)$ is,
\begin{eqnarray}
PT \varphi_{n_1}({\cal{X}}_1)=(-1)^{n_1}\frac{\sqrt{{\overline{{c}}_1}}}{\sqrt[4]{\pi}}(2^{n_1}n_1!)^{-\frac{1}{2}}e^{-\frac{1}{2}({\overline{{c}}_1}
{\cal{X}}_1)^2}H_{n_1}({\overline{{c}}_1}{\cal{X}}_1) \neq {\rm const.}\, \varphi_{n_1}({\cal{X}}_1), \label{PTc1}
\end{eqnarray}
where ${\overline{{c}}_1}$ means the complex conjugate of $c_1$. The above equation
gives rise to the result that $\varphi_{n_1n_2}({\cal{X}}_1,{\cal{X}}_2)$ is no longer the eigenfunction of $PT$.
Thus the $PT$ symmetry is now broken, which was ignored in
ref.~\cite{BJI} because the eigenfunctions were not solved there.

\section{Model 2: Two coupled non-Hermitian $PT$-symmetric Hamiltonian system}

We apply the algebraic method to a more complicated model~\cite{BJI} which is composed of
two coupled $PT$-symmetric Hamiltonians,
\begin{equation}
H=\left(p^2_1+x^2_1+i2\tau_1x_1\right)+\left(p^2_2+x^2_2+i2\tau_2x_2\right)+2\epsilon x_1 x_2,\label{PTH2}
\end{equation}
where both oscillators contain non-Hermitian terms but the interaction is Hermitian, and $\tau_1$ and $\tau_2$ are real parameters.
We shall investigate this model by following the same way as in the above section. Here we
emphasize that the analyzing procedure is almost same except for involving in more complicated calculations,
thus we give prominence to the important results but omit the related computing.

First, we write the diagonalized formulation of the above Hamiltonian eq.~(\ref{PTH2}),
\begin{equation}
H=2a^2{\cal{P}}^2_1+\frac{1+\zeta\epsilon}{2a^2}{\cal{X}}^2_1+2b^2{\cal{P}}^2_2+\frac{1-\zeta\epsilon}{2b^2}{\cal{X}}^2_2
+\frac{\tau^2_1+\tau^2_2-2\epsilon\tau_1\tau_2}{1-\epsilon^2},\label{diagH2}
\end{equation}
where the variables ${\cal{P}}_1$, ${\cal{P}}_2$ are the same as that defined in eqs.~(\ref{arguments}) and (\ref{Arguments}), but
${\cal{X}}_1$ and ${\cal{X}}_2$ take the forms,
\begin{eqnarray}
{\cal{X}}_1=a\left(x_1+\zeta x_2+\frac{i\left(\tau_1+\zeta\tau_2\right)}{1+\zeta\epsilon}\right),
&& {\cal{X}}_2=b\left(x_1-\zeta x_2+\frac{i\left(\tau_1-\zeta\tau_2\right)}{1-\zeta\epsilon}\right),\label{arguments2}
\end{eqnarray}
which are different from that of model 1 (see eq.~(\ref{Arguments})) just in the constant imaginary parts.
Comparing eq.~(\ref{diagH2}) with eq.~(\ref{comdiagH}), we see that they are almost same but have the different constant shift terms.
Therefore, the important results in the two models have to have the same formulations, such as the operator $V$ (eq.~(\ref{V})),
the new annihilation and creation
operators (eqs.~(\ref{cran}) and (\ref{cran2})), the number operator (eq.~(\ref{number})), and the associated commutation relations
(eqs.~(\ref{acr}) and (\ref{Nacr2})).

Next, we give the Hamiltonian written in terms of number operators,
\begin{eqnarray}
H=\sqrt{1+\zeta\epsilon}\left(2N_1+1\right)+\sqrt{1-\zeta\epsilon}\left(2N_2+1\right)+\frac{\tau^2_1+\tau^2_2-2\epsilon\tau_1\tau_2}{1-\epsilon^2},
\label{numberH2}
\end{eqnarray}
whose spectrum obviously has the form,
\begin{equation}
E_{n_1n_2}=\sqrt{1+\zeta\epsilon}\left(2n_1+1\right)+\sqrt{1-\zeta\epsilon}\left(2n_2+1\right)+\frac{\tau^2_1+\tau^2_2
-2\epsilon\tau_1\tau_2}{1-\epsilon^2}.
\end{equation}
In the above of this section, we focus only on the case $|\epsilon| < 1$, and consequently obtain the real and positive spectrum.
We note that this spectrum coincides with that given in ref.~\cite{BJI} where the non-Hermitian $PT$-symmetric Hamiltonian
was dealt with by being converted to its Hermitian counterpart. Although the spectrum is the same for the two different Hamiltonians,
the eigenfunctions of the non-Hermitian $PT$-symmetric Hamiltonian are different from that of the Hermitian counterpart.
In addition, the critical point is at $|\epsilon| =1$, and for the case $|\epsilon| > 1$, the spectrum becomes complex
due to the breaking of the $PT$ symmetry as analyzed in the subsection 2.5.

At last, we turn to the eigenfunctions of the Hamiltonian (see eq.~(\ref{PTH2}), eq.~(\ref{diagH2}), or  eq.~(\ref{numberH2}))
and the positive definiteness of their inner products, which was not studied in ref.~\cite{BJI}.
We can work out the same eigenfunctions as eqs.~(\ref{eigenstate}) and (\ref{state}) in which
the coordinates ${\cal{X}}_1$ and ${\cal{X}}_2$ should be replaced by the ones whose operators are
defined by eq.~(\ref{arguments2}). As to the positive definite
$PV$-pseudo inner product (or its equivalent $PTV$ inner product) of the eigenfunctions, we can prove by achieving
the similar calculations to that expressed by eq.~(\ref{pnorm}) where  the two parameters
in the upper and lower limits of integration
now take the values $I_1=\frac{a\left(\tau_1+\zeta\tau_2\right)}{1+\zeta\epsilon}$ and $I_2=\frac{b\left(\tau_1-\zeta\tau_2\right)}{1-\zeta\epsilon}$.

\section{Conclusion}

In this paper we apply the algebraic method to two non-Hermitian $PT$-symmetric quantum systems and obtain the energy spectra and
eigenfunctions, and further investigate the relation between the reality of spectra and the $PT$ symmetry of the systems.
Note that $|\epsilon| = 1$ is a critical point for the two models described by eq.~(\ref{PTH}) and eq.~(\ref{PTH2}).
In the weak interacting region, $|\epsilon| < 1$,
the spectra are real and positive and the $PT$ symmetry is unbroken; in the strong interacting region, $|\epsilon| > 1$,
the spectra are complex and the $PT$ symmetry is broken.
The spectra we obtain are exactly same as that given in ref.~\cite{BJI} where the eigenfunctions were circumvented
because the Hamiltonians of the systems were changed. Our results show that the algebraic method is available to
the non-Hermitian $PT$-symmetric quantum systems although it was proposed for the $\eta_+$-pseudo Hermitian quantum systems.
We prove the equivalence between the $PTV$ inner product and the $PV$-pseudo inner product and confirm the positive definiteness of the inner product of
eigenfunctions.
In particular, due to the reason that the operator $V$ can be constructed in terms of Hamiltonians,
our proof is free of the postulation that
a Hamiltonian should possess a complete biorthonormal eigenbasis and a discrete spectrum, while such a postulation is mandatory in
ref.~\cite{Ali3} because there the related operator was constructed in terms of eigenfunctions, see also footnote 3. As a whole,
we bring the non-Hermitian $PT$-symmetric Hamiltonian systems into the framework of the $PV$-pseudo Hermitian quantum mechanics and then
deal with them by using the algebraic method.
Comparing with the way adopted in ref.~\cite{BJI} that converts a non-Hermitian Hamiltonian to its Hermitian counterpart,
we note that our method has the merit that keeps the Hilbert space of the non-Hermitian $PT$-symmetric Hamiltonian system unchanged.

We point out that the two models (see eqs.~(\ref{PTH}) and (\ref{PTH2})) are symmetric under the transposition
and therefore they are also $P$-pseudo Hermitian
self-adjoint. However, for the models that do not have such an invariance, it is still unclear which kind of pseudo Hermitian self-adjoint symmetries
they correspond to. This is an interesting problem and thus left for our further consideration in a separate work.

\section*{Acknowledgments}
This work was supported in part by the National Natural
Science Foundation of China under grant No.11175090, and by the Fundamental Research Funds for the Central Universities under grant
No.65030021.


\baselineskip 20pt
\begin{thebibliography}{s3}
\bibitem{BBR}
   C.M. Bender and S. Boettcher, {\em Real spectra in non-Hermitian Hamiltonians having $PT$ symmetry},
   Phys. Rev. Lett. {\bf 80} (1998) 5243 [arXiv:math-ph/9712001].

\bibitem{BBMP}
    C.M. Bender, S. Boettcher and P.N. Meisinger, {\em $PT$-symmetric quantum mechanics}, J. Math. Phys. {\bf 40} (1999) 2201 [arXiv:quant-ph/9809072].

\bibitem{BQZG}
   B. Bagchi, C. Quesne and M. Znojil, {\em Generalized continuity equation and modified normalization in $PT$-symmetric quantum mechanics},
   Mod. Phys. Lett.  A {\bf 16} (2001) 2047 [arXiv:quant-ph/0108096].

\bibitem{BBJM}
    C.M. Bender, D.C. Brody and H.F. Jones, {\em Complex extension of quantum mechanics}, Phys. Rev. Lett. {\bf 89} (2002) 270401
    [Erratum: Ibid. {\bf 92} (2004) 119902] [arXiv:quant-ph/0208076].

\bibitem{SW}
    S. Weigert, {\em Completeness and orthonormality in $PT$-symmetric quantum systems}, Phys. Rev. A {\bf 68} (2003) 062111 [arXiv:quant-ph/0306040].

\bibitem{LF}
L. Feng, et al., {\em Nonreciprocal light propagation in a silicon photonic circuit}, Science {\bf 333} (2011) 729 [arXiv:]; and the references therein.


\bibitem{Pauli}
    W. Pauli, {\em On Dirac's new method of field quantization}, Rev. Mod. Phys. {\bf 15} (1943) 175.

\bibitem{SGH}
F.G. Scholtz, H.B. Geyer and F.J.W. Hahne, {\em Quasi-Hermitian operators in quantum mechanics and the variational principle},
Ann. Phys. (N.Y.) {\bf 213} (1992) 74.

\bibitem{Ali1}
    A. Mostafazadeh, {\em Pseudo-Hermiticity versus $PT$ symmetry: The necessary condition for the reality of the spectrum of a
    non-Hermitian Hamiltonian}, J. Math. Phys. {\bf 43} (2002) 205 [arXiv:math-ph/0107001].

\bibitem{Ali2}
    A. Mostafazadeh, {\em Pseudo-Hermiticity versus $PT$-Symmetry III: Equivalence of pseudo-Hermiticity and the presence of antilinear symmetries},
    J. Math. Phys. {\bf 43} (2002) 3944 [arXiv:math-ph/0203005].

\bibitem{Ali3}A. Mostafazadeh,
{\em Pseudo-Hermiticity and generalized PT and CPT symmetries},  J. Math. Phys. {\bf 44} (2003) 974 [arXiv:math-ph/0209018].


\bibitem{ME}
    A. Mostafazadeh, {\em Exact $PT$-symmetry is equivalent to Hermiticity}, J. Phys. A {\bf 36} (2003) 7081 [arXiv:quant-ph/0304080].

\bibitem{BMN}
A. Pais and G.E. Uhlenbeck, {\em On field theories with non-localized action}, Phys. Rev. {\bf 79} (1950) 145;\\
    C.M. Bender and P.D. Mannheim, {\em No-ghost theorem for the fourth-order derivative Pais-Uhlenbeck oscillator model},
    Phys. Rev. Lett. {\bf 100} (2008) 110402 [arXiv:0706.0207 [hep-th]].

%\bibitem{BBJMF}
%   C.M. Bender, D.C. Brody, H.F. Jones and B.K. Meister, {\em Faster than Hermitian quantum mechanics}, SIGMA {\bf 3} (2007) 126
%   [arXiv:0712.3910 [hep-th]].

\bibitem{BJI}
    C.M Bender and H.F Jones, {\em Interactions of Hermitian and non-Hermitian Hamiltonians}, J. Phys. A {\bf 41} (2008) 244006
    [arXiv:0709.3605 [hep-th]].

\bibitem{LMX}J.-Q. Li, Y.-G. Miao and  Z. Xue, {\em Algebraic Method for Pseudo-Hermitian Hamiltonian}, arXiv:1107.4972[quant-ph];\\
J.-Q. Li and  Y.-G. Miao, {\em Spontaneous Breaking of Permutation Symmetry in Pseudo-Hermitian Quantum Mechanics}, Phys. Rev. A
{\bf 85} (2012) 042110 [arXiv:1110.2312[quant-ph]].

%\bibitem{MBW}
%    G.A. Mezincescu, {\em Some properties of eigenvalues and eigenfunctions of the cubic
%    oscillator with imaginary coupling constant}, J. Phys. A {\bf 33} (2000) 4911 [arXiv:quant-ph/0002056];\\
%    C.M. Bender and Q. Wang, {\em Comment on a recent paper by Mezincescu}, J. Phys. A {\bf 34} (2001) 3325.
\end{thebibliography}
\end{document}